\documentclass[11pt,a4paper]{article}

\usepackage[utf8]{inputenc}
\usepackage[T1]{fontenc}
\usepackage{lmodern}
\usepackage[margin=2.5cm]{geometry}
\usepackage{amsmath,amssymb}
\usepackage{graphicx}
\usepackage{booktabs}
\usepackage{xspace}
\usepackage{authblk}
\usepackage[numbers,sort&compress]{natbib}
\usepackage{tikz}
\usetikzlibrary{arrows.meta}
\usepackage[colorlinks=true,linkcolor=blue,citecolor=blue,urlcolor=blue]{hyperref}

\newcommand{\kk}{\kappa}                 
\newcommand{\muo}{\mu_0}
\newcommand{\unit}[1]{\,\mathrm{#1}}
\newcommand{\cfg}[1]{\textsf{#1}}        
\newcommand{\etal}{\emph{et al.}\xspace}

\newcommand{\magnet}[3]{%
  \draw[thick,fill=blue!6] (#1-0.5,#2-0.5) rectangle (#1+0.5,#2+0.5);
  \draw[-{Stealth[length=2.2mm,width=1.8mm]},red,line width=1.1pt]
       (#1,#2-0.33*#3) -- (#1,#2+0.33*#3);
}
\newcommand{\sensor}[3]{
  \fill[cyan!35] (#1,#2) circle (0.16);
  \draw[thick,cyan!60!black] (#1,#2) circle (0.16);
}
\newcommand{\Bvec}[4]{%
  \draw[-{Stealth[length=2.4mm,width=2mm]},blue!70!black,line width=1.2pt]
       (#1,#2) -- (#1+#3,#2+#4);
}

\title{Experimental characterization of magnetic multipole fields\\ up to octupole order using smartphone magnetometry}

\author[1]{Mart\'{\i}n Monteiro}
\author[2]{Cecilia Stari}
\author[2,*]{Arturo C. Mart\'{\i}}
\author[3]{Enrique Arribas}
\author[3]{Isabel Escobar}
\author[3]{Raquel Ramirez-Vazquez}
\author[4]{Augusto Bel\'endez}

\affil[1]{Universidad ORT Uruguay, Montevideo, Uruguay}
\affil[2]{Instituto de F\'{\i}sica, Facultad de Ciencias, Universidad de la Rep\'ublica, Igu\'a 4225, 11400 Montevideo, Uruguay}
\affil[3]{Applied Physics Department, University of Castilla-La Mancha, 02071 Albacete, Spain}
\affil[4]{Departamento de F\'{\i}sica, Ingenier\'{\i}a de Sistemas y Teor\'{\i}a de la Se\~nal, Universidad de Alicante, 03080 Alicante, Spain}
\affil[*]{Corresponding author: \texttt{marti@fisica.edu.uy}}

\date{\today}

\begin{document}
\maketitle

\begin{abstract}
The multipole expansion is one of the organising ideas of electromagnetism: far from a localised source the field is dominated by its lowest non-vanishing moment, and each successive order decays one power of the distance faster. Although the dipole field is routinely measured in introductory laboratories, higher orders are seldom explored experimentally, because the fields are weak, decay very rapidly and require careful positioning. Here we show that with four identical neodymium cube magnets, a sheet of graph paper and the magnetometer of a smartphone, first-year students can build and characterise not one but three distinct magnetic octupoles (linear, planar and `square' arrangements), together with the dipole and the two quadrupoles that serve as reference. A simple notation and a compact set of formulae, based on the spatial Cartesian moments $M_{pq}=\sum_k m_k x_k^p z_k^q$ of the array, predict which field component survives at each measuring position and how it decays. The measured exponents are $n\simeq 3$, $4$ and $5$ for dipole, quadrupole and octupole arrangements, and the measured prefactors show fair quantitative consistency with the independently determined dipole moment of the magnets once nominal geometry and spacer effects are taken into account. We discuss the systematic effects that limit the octupole measurements (higher-order terms at short distance, residual lower-order moments at long distance, the location of the sensor inside the phone) and the ways in which the activity can be extended. The experiment generalises and unifies previous smartphone-based measurements of dipoles, linear quadrupoles and a linear octupole.
\end{abstract}


\section{Introduction}
\label{sec:intro}

The multipole expansion expresses the electric or magnetic field of a localised source as a series in inverse powers of the distance $r$ to the source \citep{Jackson1999,Griffiths2017,Raab2004}. Each term is characterised by a multipole moment (monopole, dipole, quadrupole, octupole, \dots) and by a decay law: an electric monopole field falls off as $1/r^2$, a dipole field as $1/r^3$, a quadrupole field as $1/r^4$, an octupole field as $1/r^5$, and so on. Since magnetic monopoles do not exist, the leading term of any magnetostatic source is the dipole, and arrangements of magnets in which the dipole moment cancels expose the quadrupole, octupole and higher terms \citep{Castellanos1978}. The expansion is far more than a mathematical convenience. It is the language in which the main field of the Earth is described---the International Geomagnetic Reference Field is a spherical-harmonic expansion whose leading terms are the geocentric dipole, quadrupole and octupole \citep{IGRF13}---and it is used in the same way for planets and stars. In atomic and nuclear physics the hyperfine structure of atomic levels is parametrised by the nuclear magnetic dipole, electric quadrupole and magnetic octupole moments \citep{Schwartz1955}, and the measurement of nuclear magnetic octupole moments in alkali atoms is an active field of research \citep{Gerginov2003}. In chemistry the long-range part of intermolecular forces is organised as a multipole series \citep{Stone2013}, and in condensed-matter physics higher-order magnetic multipoles serve as hidden order parameters in $f$-electron compounds and antiferromagnets \citep{Santini2009}. On the engineering side, the magnets that guide and focus particle beams in accelerators are called dipoles, quadrupoles, sextupoles and octupoles after the symmetry of the field they produce \citep{Wiedemann2015,Sullivan2023}; the shimming of magnetic-resonance magnets consists in cancelling, order by order, the unwanted multipole components of the field \citep{Romeo1984}; and permanent-magnet assemblies are designed to produce pure multipole fields for atom and ion traps \citep{Kaenders1993}.

From a conceptual point of view the multipole expansion teaches students two complementary lessons. The first is a cancellation hierarchy: two equal and opposite dipoles placed close together have no dipole moment, so that their field is dominated by the next term; two such quadrupoles placed in opposition have neither dipole nor quadrupole moment, and so on. The second is that symmetry alone decides which components of the field survive at a given point, before any calculation is made. Both lessons are best learned by doing. However, while the inverse-cube law of a small magnet is a standard introductory experiment, the quadrupole and especially the octupole are rarely measured in first-year laboratories: their fields are weak, they decay very fast, and traditional Hall probes and gaussmeters are expensive and require careful alignment. As Sullivan \etal\ remark, examples and explanations beyond the dipole are scarce \citep{Sullivan2023}.

Smartphones offer a simple, accurate and inexpensive way out of this difficulty. Many current smartphones contain a three-axis magnetometer with a resolution of a fraction of a microtesla and a range of a few millitesla, and free applications give direct access to the readings \citep{Monteiro2022,Staacks2018}. The use of these sensors in physics teaching is now well documented---see the Resource Letter \citep{Monteiro2022} and references therein---and magnetostatics has been one of the most fruitful areas: the inverse-cube law of small magnets \citep{Arribas2015,Lara2014,Silva2012}, the field of coils and the permeability of vacuum \citep{Wannous2023}, the field of electric railways \citep{Monteiro2020}, `fly-by' measurements combining magnetometer and accelerometer \citep{Monteiro2017}, the field of a linear quadrupole \citep{Arribas2020}, small-scale analogues of accelerator magnets \citep{Sullivan2023} and, more recently, a linear octupole built with four ring magnets \citep{Arribas2024}.

Beyond the educational environment, the rapid and quantitative characterization of complex magnetic configurations using consumer electronics also offers a pragmatic approach for low-cost instrumentation. In applied physics and engineering, the ability to map multipole fields and verify theoretical decay laws without requiring expensive commercial gaussmeters or cryogenic Hall probes provides a versatile tool for the rapid prototyping of permanent-magnet arrays, sensor calibration, and preliminary benchtop validation of magnetic assemblies.

In this work we extend these previous studies and place them in a general framework. Starting from four identical cube magnets we show that there is not one but three essentially different ways of building an octupole (two linear arrangements, with the moments along or perpendicular to the line of magnets, and a planar $2\times 2$ arrangement), that each of them can be probed at two positions (on the magnetisation axis or in the equatorial plane) and that at each position a different field component survives. A compact notation and a single set of formulae, written in terms of the Cartesian moments of the array, cover all cases, including the dipole and the two quadrupoles that we measure as reference. Section~\ref{sec:exp} presents the configurations, the theory and the experimental procedure; section~\ref{sec:results} the measurements; section~\ref{sec:discussion} discusses the results and their limitations; and section~\ref{sec:conclusions} summarises the conclusions and suggests extensions.

\section{The experiment}
\label{sec:exp}

\subsection{Building multipoles with identical magnets}
\label{sec:config}

\begin{figure}[tbp]
\centering
\begin{tikzpicture}[scale=0.56]
\useasboundingbox (-2.9,-4.3) rectangle (5.8,5.6);
\node[anchor=north west] at (-2.9,5.6) {(a)};
\draw[->] (-1.6,0) -- (5.4,0) node[below] {$x$};
\draw[->] (0,-2.6) -- (0,5.3) node[left] {$z$};
\magnet{0}{0}{1}
\sensor{0}{3.2}{}    \node[right=3pt] at (0.16,3.2) {\cfg{Dzp}};
\Bvec{0}{3.2}{0}{0.9}  \node[right=2pt,blue!70!black] at (0.1,4.1) {$\mathbf B$};
\sensor{3.6}{0}{}    \node[above=3pt] at (3.6,0.16) {\cfg{Drt}};
\Bvec{3.6}{0}{0}{-0.9} \node[right=2pt,blue!70!black] at (3.7,-0.9) {$\mathbf B$};
\end{tikzpicture}\hfill
\begin{tikzpicture}[scale=0.56]
\useasboundingbox (-2.9,-4.3) rectangle (5.8,5.6);
\node[anchor=north west] at (-2.9,5.6) {(b)};
\draw[->] (-1.6,0) -- (5.4,0) node[below] {$x$};
\draw[->] (0,-2.6) -- (0,5.3) node[left] {$z$};
\magnet{0}{1.0}{1}
\magnet{0}{-1.0}{-1}
\draw[<->,gray] (1.0,1.0) -- (1.0,-1.0) node[midway,right,black] {$d$};
\sensor{0}{3.2}{}    \node[right=3pt] at (0.16,3.2) {\cfg{QLzp}};
\Bvec{0}{3.2}{0}{0.9}  \node[right=2pt,blue!70!black] at (0.1,4.1) {$\mathbf B$};
\sensor{3.6}{0}{}    \node[above=3pt] at (3.6,0.16) {\cfg{QLrp}};
\Bvec{3.6}{0}{-0.9}{0} \node[below=2pt,blue!70!black] at (2.8,-0.1) {$\mathbf B$};
\end{tikzpicture}\hfill
\begin{tikzpicture}[scale=0.56]
\useasboundingbox (-2.9,-4.3) rectangle (5.8,5.6);
\node[anchor=north west] at (-2.9,5.6) {(c)};
\draw[->] (-1.6,0) -- (5.4,0) node[below] {$x$};
\draw[->] (0,-2.6) -- (0,5.3) node[left] {$z$};
\magnet{-0.5}{0}{-1}
\magnet{0.5}{0}{1}
\sensor{0}{3.2}{}    \node[right=3pt] at (0.16,3.2) {\cfg{QPzt}};
\Bvec{0}{3.2}{-0.9}{0} \node[above=2pt,blue!70!black] at (-0.8,3.3) {$\mathbf B$};
\sensor{3.6}{0}{}    \node[above=3pt] at (3.6,0.16) {\cfg{QPrt}};
\Bvec{3.6}{0}{0}{-0.9} \node[right=2pt,blue!70!black] at (3.7,-0.9) {$\mathbf B$};
\end{tikzpicture}

\vspace{2mm}

\begin{tikzpicture}[scale=0.56]
\useasboundingbox (-2.9,-4.3) rectangle (5.8,5.6);
\node[anchor=north west] at (-2.9,5.6) {(d)};
\draw[->] (-1.6,0) -- (5.4,0) node[below] {$x$};
\draw[->] (0,-3.4) -- (0,5.3) node[left] {$z$};
\magnet{0}{2.3}{1}
\magnet{0}{0.5}{-1}
\magnet{0}{-0.5}{-1}
\magnet{0}{-2.3}{1}
\draw[<->,gray] (1.0,0) -- (1.0,0.5) node[midway,right,black] {\footnotesize $z_1$};
\draw[<->,gray] (1.6,0) -- (1.6,2.3) node[midway,right,black] {\footnotesize $z_2$};
\sensor{0}{3.4}{}    \node[right=3pt] at (0.16,3.4) {\cfg{OLLzp}};
\Bvec{0}{3.4}{0}{0.9}  \node[right=2pt,blue!70!black] at (0.1,4.3) {$\mathbf B$};
\end{tikzpicture}\hfill
\begin{tikzpicture}[scale=0.56]
\useasboundingbox (-2.9,-4.3) rectangle (5.8,5.6);
\node[anchor=north west] at (-2.9,5.6) {(e)};
\draw[->] (-1.6,0) -- (5.4,0) node[below] {$x$};
\draw[->] (0,-2.6) -- (0,5.3) node[left] {$z$};
\magnet{-0.5}{1.0}{-1}
\magnet{0.5}{1.0}{1}
\magnet{-0.5}{-1.0}{1}
\magnet{0.5}{-1.0}{-1}
\draw[<->,gray] (1.5,0) -- (1.5,1.0) node[midway,right,black] {\footnotesize $z_1$};
\draw[<->,gray] (0,-2.0) -- (0.5,-2.0) node[right,black] {\footnotesize $x_1$};
\sensor{0}{3.2}{}    \node[right=3pt] at (0.16,3.2) {\cfg{OPLzt}};
\Bvec{0}{3.2}{-0.9}{0} \node[above=2pt,blue!70!black] at (-0.8,3.3) {$\mathbf B$};
\sensor{3.6}{0}{}    \node[above=3pt] at (3.6,0.16) {\cfg{OPLrp}};
\Bvec{3.6}{0}{-0.9}{0} \node[below=2pt,blue!70!black] at (2.8,-0.1) {$\mathbf B$};
\end{tikzpicture}\hfill
\begin{tikzpicture}[scale=0.56]
\useasboundingbox (-2.8,-4.6) rectangle (5.6,5.6);
\node[anchor=north west] at (-2.8,5.6) {(f)};
\draw[->] (-2.8,0) -- (5.6,0) node[below] {$x$};
\draw[->] (0,-2.6) -- (0,5.3) node[left] {$z$};

\magnet{-2.0}{0}{1}
\magnet{-1.0}{0}{-1}
\magnet{1.0}{0}{-1}
\magnet{2.0}{0}{1}

\draw[dashed,fill=gray!15] (-0.5,-0.5) rectangle (0.5,0.5);
\draw[<->,gray!80!black] (-0.5,0.7) -- (0.5,0.7) node[midway,above,black] {\footnotesize $s$};

\draw[dotted,gray] (1.0,-0.5) -- (1.0,-2.6);
\draw[dotted,gray] (2.0,-0.5) -- (2.0,-2.6);
\draw[dotted,gray] (0,-0.5) -- (0,-2.6);

\draw[<->,black] (0,-1.3) -- (1.0,-1.3) node[midway,fill=white,inner sep=1pt] {\footnotesize $x_1$};
\draw[<->,black] (0,-2.3) -- (2.0,-2.3) node[midway,fill=white,inner sep=1pt] {\footnotesize $x_2$};

\sensor{0}{3.2}{}    \node[right=3pt] at (0.16,3.2) {\cfg{OPPzp}};
\Bvec{0}{3.2}{0}{-0.9} \node[right=2pt,blue!70!black] at (0.1,2.3) {$\mathbf B$};
\sensor{3.9}{0}{}    \node[above=3pt] at (3.9,0.16) {\cfg{OPPrt}};
\Bvec{3.9}{0}{0}{-0.9} \node[right=2pt,blue!70!black] at (4.0,-0.9) {$\mathbf B$};
\end{tikzpicture}
\caption{The configurations studied. Each square is a cube magnet of edge $a$ and the red arrow is its magnetic moment; the small circles mark the two positions of the magnetometer, on the $z$ axis (axial, \cfg{z}) or on the $x$ axis in the equatorial plane (radial, \cfg{r}), and the blue arrow at each sensor shows the direction of the field component that is measured there: along the line joining the centre of the array to the sensor for the polar component (\cfg{p}), perpendicular to it for the tangential component (\cfg{t}). The labels give the code of the corresponding measurement. (a) Dipole. (b) Longitudinal quadrupole \cfg{QL}: two magnets stacked along their axis with a spacer, centre-to-centre distance $d$. (c) Parallel quadrupole \cfg{QP}: two magnets side by side. (d) Linear octupole \cfg{OLL}: two \cfg{QL} quadrupoles in opposition, magnets centred at $\pm z_1=\pm a/2$ and $\pm z_2$. (e) Planar octupole \cfg{OPL}: two \cfg{QP} quadrupoles stacked in opposition (or two \cfg{QL} side by side), magnets centred at $(\pm x_1,\pm z_1)$ with $x_1=a/2$ and $z_1=(a+s)/2$. (f) Linear octupole \cfg{OPP}: two \cfg{QP} quadrupoles side by side in opposition, magnets centred at $\pm x_1=\pm(a+s)/2$ and $\pm x_2=\pm(3a+s)/2$, separated by a central gap of thickness $s$. All positional coordinates ($x_k, z_k$) are referenced from the array centre to the centres of the magnets.}
\label{fig:config}
\end{figure}
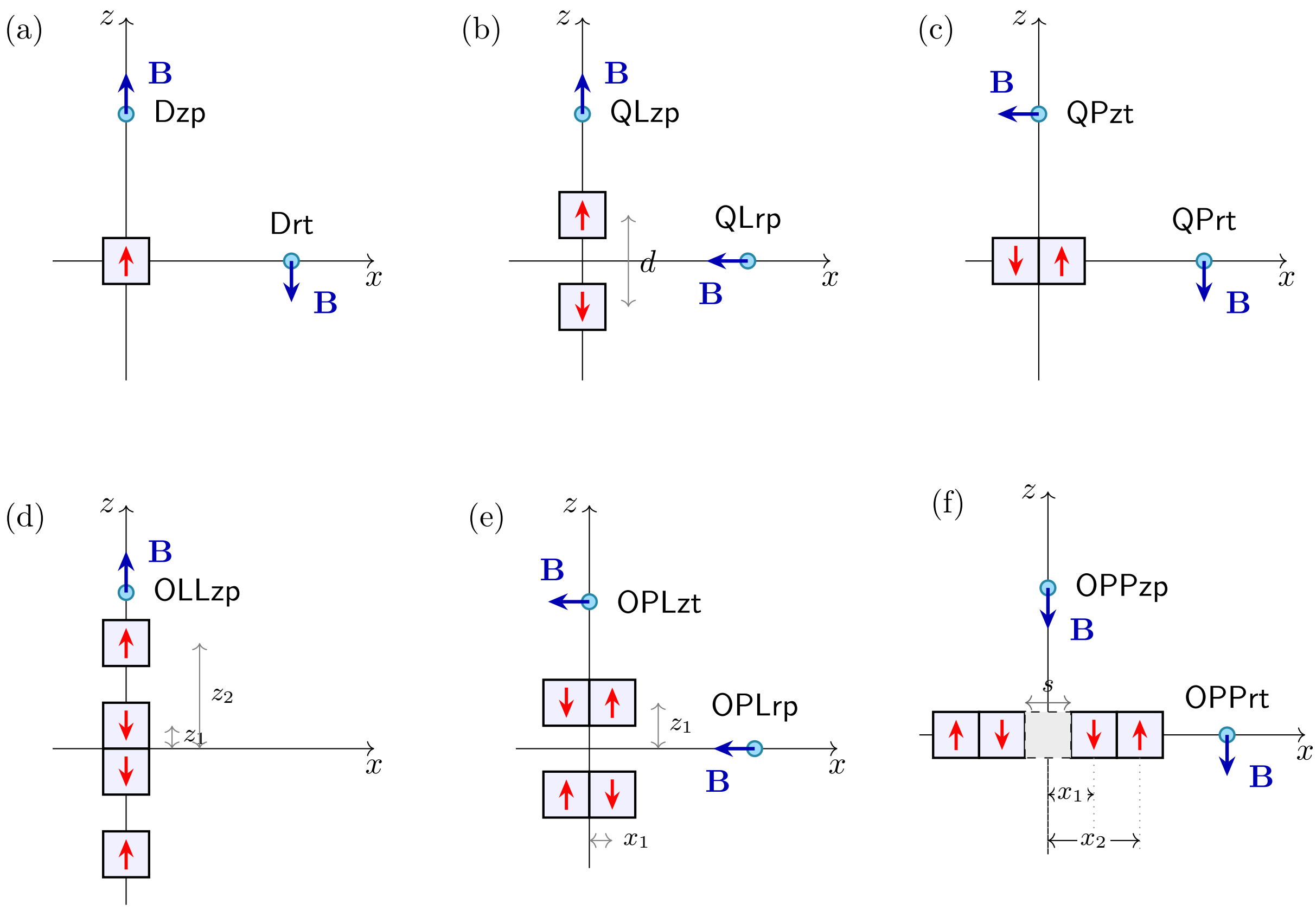

All the arrangements studied here are built from identical cube magnets of edge $a$ and magnetic moment of magnitude $m$. We take the magnetisation axis of the magnets as the $z$ axis, so that each magnet $k$ carries a moment $m_k\hat{\mathbf z}$ with $m_k=\pm m$. Radial measurements are performed along the $x$ axis in the equatorial plane, with coordinates $(x,0,0)$ where $x>0$ (figure~\ref{fig:config}). Two magnets with opposite moments can be joined in two ways: stacked along their common axis, one above the other (we call this a \emph{longitudinal} quadrupole, \cfg{QL}), or side by side, with their axes parallel (a \emph{parallel} quadrupole, \cfg{QP}). The first is the linear quadrupole of \citet{Arribas2020}. Note that in \cfg{QP} the magnets attract each other and the arrangement is self-supporting, whereas in \cfg{QL} they repel and have to be held apart by a non-magnetic spacer; the centre-to-centre distance in \cfg{QL} will be called $d$.

Octupoles are obtained by joining two quadrupoles in opposition, so that the quadrupole moments cancel as well. Again there are several possibilities, which gives rise to the three-letter notation of figure~\ref{fig:config}:
\begin{itemize}
\item \cfg{OLL}: two longitudinal quadrupoles stacked along $z$ with opposite orientation, i.e.\ four collinear magnets with moments along the line in the sequence $+\,-\,-\,+$. The inner magnets, at $z=\pm z_1$ with $z_1=a/2$, attract each other; the outer ones, at $z=\pm z_2$, must be held by spacers. This is the linear octupole of \citet{Arribas2024}, there built with ring magnets.
\item \cfg{OPP}: two parallel quadrupoles placed side by side along $x$, i.e.\ four collinear magnets with moments \emph{perpendicular} to the line, in the sequence $+\,-\,-\,+$. The outer magnets are at $x=\pm x_2$ and the inner ones at $x=\pm x_1$; the two inner magnets repel, so that a central spacer of thickness $s$ is placed between the two quadrupoles.
\item \cfg{OPL} (equivalently \cfg{OLP}): two parallel quadrupoles stacked along $z$ with opposite orientation, which is the same thing as two longitudinal quadrupoles placed side by side; the four magnets occupy the corners $(\pm x_1,\pm z_1)$ of a rectangle, with $x_1=a/2$, and the sign of each moment is the product of the signs of its coordinates. The two rows repel and are separated by a spacer of thickness $s$, giving $z_1=(a+s)/2$.
\end{itemize}
Students readily find a fourth candidate, the alternating sequence $+\,-\,+\,-$ of four magnets side by side. We shall see below that it is \emph{not} an octupole: it has a non-vanishing quadrupole moment and its field decays as $1/x^4$.

Each arrangement is probed with the magnetometer placed either on the $z$ axis (the magnetisation axis; we call this the \emph{axial} position, letter \cfg{z}) or on the $x$ axis in the equatorial plane (\emph{radial} position, letter \cfg{r}). At each position the field may have a component along the line joining the centre of the array to the sensor, which we call the \emph{polar} component (letter \cfg{p}), and a component perpendicular to that line, the \emph{tangential} component (letter \cfg{t}); the blue arrows in figure~\ref{fig:config} show, for each configuration and position, the component that survives and is measured. For a single magnet the axial and radial positions are the classical `end-on' and `broadside' (Gauss) positions, in which the field is polar and tangential, respectively. The code \cfg{OPPrt}, for instance, denotes the tangential component of the \cfg{OPP} octupole measured on the radial axis. For a single magnet the two measurements are \cfg{Dzp} and \cfg{Drt}; the complete set of configurations studied here is listed in table~\ref{tab:moments}.

\subsection{Multipole expansion for an array of parallel dipoles}
\label{sec:theory}

At distances large compared with the overall size of the array, each cube magnet is well described as a point dipole \citep{Arribas2015,Binder2016}, whose field is
\begin{equation}
\mathbf B(\mathbf R)=\kk\,\frac{3(\mathbf m\cdot\hat{\mathbf R})\hat{\mathbf R}-\mathbf m}{R^3},
\qquad \kk=\frac{\muo}{4\pi}=10^{-7}\unit{T\,m\,A^{-1}},
\label{eq:dipole}
\end{equation}
where $\mathbf R$ is the vector from the magnet to the field point. For a single magnet, with $\mathbf m=m\hat{\mathbf z}$, equation~(\ref{eq:dipole}) gives the two classical results: on the axis the field is polar, $B_z=2\kk m/z^3$ (\cfg{Dzp}), and in the equatorial plane it is tangential, $B_z=-\kk m/x^3$ (\cfg{Drt}), half as large and antiparallel to $\mathbf m$.

For an array of magnets the field is the superposition of the individual dipole fields, and the multipole expansion is the expansion of that sum in inverse powers of the observation distance. For the arrangements of figure~\ref{fig:config} all the magnets lie in the $xz$ plane, at positions $(x_k,0,z_k)$, and all the moments are along $z$. It is then convenient to define the Cartesian moments of the array,
\begin{equation}
M_{pq}=\sum_k m_k\,x_k^{\,p}\,z_k^{\,q},
\label{eq:Mpq}
\end{equation}
where $p+q$ is the degree of the spatial moment (corresponding to a magnetic multipole order $\ell = p+q+1$): $M_{00}$ is the total dipole moment ($\ell=1$), $M_{10}$ and $M_{01}$ are the quadrupole moments ($\ell=2$) that can be built with parallel dipoles, and $M_{20}$, $M_{11}$ and $M_{02}$ are the octupole moments ($\ell=3$). These are the Cartesian moments of the magnetisation distribution $\mathbf M(\mathbf r)=\sum_k m_k\hat{\mathbf z}\,\delta(\mathbf r-\mathbf r_k)$, i.e.\ the non-vanishing components of the tensors $\int\mathbf M\,\mathbf r\cdots\mathbf r\,{\rm d}V$ of the Cartesian multipole expansion of magnetostatics \citep{Castellanos1978,Raab2004}, and the magnetic analogue of the moments $\sum_i q_i x_i^{\,p}z_i^{\,q}$ of a discrete charge distribution \citep{Griffiths2017}. The values of these moments for our configurations are collected in table~\ref{tab:moments}. Note in particular that the alternating sequence $+\,-\,+\,-$, with the magnets at $x=\pm a/2,\pm 3a/2$, has $M_{10}=-2ma\neq0$ and is therefore a quadrupole, while the $+\,-\,-\,+$ sequence has $M_{10}=0$ and is a genuine octupole.

Expanding equation~(\ref{eq:dipole}) for a field point on the $z$ axis at a distance $z$ large compared with the source dimensions and summing over the magnets one finds, up to the octupole order,
\begin{align}
B_z(z)&=\kk\left[\frac{2M_{00}}{z^3}+\frac{6M_{01}}{z^4}+\frac{12M_{02}-6M_{20}}{z^5}+\cdots\right],
\label{eq:axial_p}\\
B_x(z)&=-3\kk\left[\frac{M_{10}}{z^4}+\frac{4M_{11}}{z^5}+\cdots\right],
\label{eq:axial_t}
\end{align}
for the polar and tangential components, respectively. For a field point in the equatorial plane along the $x$ axis at $(x,0,0)$ with $x$ large compared with the array size,
\begin{align}
B_z(x)&=-\kk\left[\frac{M_{00}}{x^3}+\frac{3M_{10}}{x^4}+\frac{6M_{20}-\frac{9}{2}M_{02}}{x^5}+\cdots\right],
\label{eq:radial_t}\\
B_x(x)&=-3\kk\left[\frac{M_{01}}{x^4}+\frac{4M_{11}}{x^5}+\cdots\right],
\label{eq:radial_p}
\end{align}
for the tangential and polar components. The derivation is elementary: for collinear arrangements along $z$ the sum reduces to a geometric-type series, $\sum_k m_k (z-z_k)^{-3}=z^{-3}\sum_k m_k[1+3z_k/z+6z_k^2/z^2+\cdots]$, and the same binomial expansion handles transverse displacements $x_k$ once equation~(\ref{eq:dipole}) is written in components. Because the non-collinear arrangements (\cfg{QP}, \cfg{OPP}, \cfg{OPL}) lack full rotational symmetry around $z$, equations~(\ref{eq:radial_t}) and (\ref{eq:radial_p}) specifically describe the field evaluated along the symmetry line $y=0$ in the magnet plane.

Equations~(\ref{eq:axial_p})--(\ref{eq:radial_p}) contain, in compact form, everything that is needed to design and interpret the experiment. (i) The decay law is fixed by the order of the lowest non-vanishing moment: $1/x^3$ for a dipole, $1/x^4$ for a quadrupole, $1/x^5$ for an octupole (using $x$ to denote the general distance from the array centre). (ii) Symmetry decides which component survives: on the $z$ axis the moments $M_{00}$, $M_{01}$, $M_{02}$ and $M_{20}$ feed the polar component while $M_{10}$ and $M_{11}$ feed the tangential one; on the equatorial $x$ axis $M_{00}$, $M_{10}$, $M_{20}$ and $M_{02}$ feed the tangential component and $M_{01}$, $M_{11}$ feed the polar one. Thus the same \cfg{OPL} array produces a purely tangential field on its axis (\cfg{OPLzt}) and a purely polar field in its equatorial plane (\cfg{OPLrp}), and the other components vanish identically at the point-dipole level. (iii) To unify the analysis across all configurations, we express the leading term of the field magnitude as $|B|=f\kk m a^{n-3}/x^n$ (with $n=3, 4, 5$), where $f$ is a dimensionless geometric prefactor that depends on the relative positions of the magnets and the presence of spacers, as listed in table~\ref{tab:moments}. Note that the polar-to-tangential ratio is 2 for the dipole and for the \cfg{QL} quadrupole, but 1 for the \cfg{QP} quadrupole and for the \cfg{OPP} and \cfg{OPL} octupoles.

\begin{table}[tbp]
\centering
\caption{Configurations studied (see figure~\ref{fig:config}), lowest non-vanishing Cartesian moment (equation~(\ref{eq:Mpq})), measuring position, surviving component, leading term of the field magnitude according to equations~(\ref{eq:axial_p})--(\ref{eq:radial_p}) ($\kk=\muo/4\pi$), and dimensionless geometric prefactor $f$ writing the leading term as $|B|=f\kk ma^{n-3}/x^n$. In \cfg{OLL}, \cfg{OPL} and \cfg{OPP} the inner magnets are at $z_1=a/2$ or $x_1=a/2$ when touching; a spacer of thickness $s$ between repelling magnets adds $s$ to the corresponding separation. For arrangements with repelling magnets, the two nominal values of $f$ correspond to touching magnets ($s=0$) and to a spacer $s=a$ (with the value used in table~\ref{tab:results} shown in parentheses when a different experimental spacer was employed).}
\label{tab:moments}
\vspace{0.5ex}
\footnotesize
\setlength{\tabcolsep}{3.5pt}
\begin{tabular}{@{}llllll@{}}
\toprule
Code & Lowest moment & Pos. & Comp. & $|B|$ (leading term) & $f$ ($s=0\mid a$)\\
\midrule
\cfg{Dzp}   & $M_{00}=m$ & axial  & polar & $2\kk m/z^3$ & $2$\\
\cfg{Drt}   & $M_{00}=m$ & radial & tang. & $\kk m/x^3$ & $1$\\
\cfg{QLzp}  & $M_{01}=md$ & axial & polar & $6\kk md/z^4$ & $6\mid 12$\\
\cfg{QLrp}  & $M_{01}=md$ & radial & polar & $3\kk md/x^4$ & $3\mid 6$\\
\cfg{QPzt}  & $M_{10}=ma$ & axial & tang. & $3\kk ma/z^4$ & $3$\\
\cfg{QPrt}  & $M_{10}=ma$ & radial & tang. & $3\kk ma/x^4$ & $3$\\
\cfg{OLLzp} & $M_{02}=2m(z_2^2-z_1^2)$ & axial & polar & $24\kk m(z_2^2-z_1^2)/z^5$ & $48\mid 144\ (192^{\rm a})$\\
\cfg{OPPzp} & $M_{20}=2m(x_2^2-x_1^2)$ & axial & polar & $12\kk m(x_2^2-x_1^2)/z^5$ & $24\mid 36$\\
\cfg{OPPrt} & $M_{20}=2m(x_2^2-x_1^2)$ & radial & tang. & $12\kk m(x_2^2-x_1^2)/x^5$ & $24\mid 36\ (48^{\rm b})$\\
\cfg{OPLzt} & $M_{11}=4mx_1z_1$ & axial & tang. & $48\kk mx_1z_1/z^5$ & $12\mid 24\ (24^{\rm c})$\\
\cfg{OPLrp} & $M_{11}=4mx_1z_1$ & radial & polar & $48\kk mx_1z_1/x^5$ & $12\mid 24\ (24^{\rm c})$\\
\bottomrule
\end{tabular}
\par\smallskip
\begin{minipage}{\linewidth}\footnotesize\raggedright
$^{\rm a}$ Value for $z_2^2-z_1^2=8a^2$ (\cfg{OLLzp}), corresponding to $z_1=a/2$ and $z_2\simeq 3a/2+s$ with $s = (\sqrt{33}-3)a/2 \simeq 1.37a$ ($\simeq 8.7\unit{mm}$).\par
$^{\rm b}$ Value for $x_2^2-x_1^2=4a^2$ (\cfg{OPPrt}), corresponding to $x_1=(a+s)/2$ and $x_2=(3a+s)/2$ with $s=2a$ ($12.7\unit{mm}$).\par
$^{\rm c}$ Value for $x_1=a/2$ and $z_1=a$, corresponding to nominal spacer $s=a$ ($6.35\unit{mm}$).
\end{minipage}
\end{table}

\subsection{Experimental procedure}
\label{sec:procedure}

The magnets are $1/4$-inch ($a=6.35\unit{mm}$) NdFeB cubes, nominally identical, from the same batch. Their dipole moment was determined independently from the period of oscillation of a bar made of the four cubes stacked with parallel moments (length $4a$, mass $7.6(1)\unit{g}$) in the horizontal component of the Earth's field, $B_{\rm h}=21.5(3)\unit{\mu T}$, measured with the same smartphone. From the period $T=0.89(1)\unit{s}$ and the moment of inertia $I=M(L^2+a^2)/12$ one obtains $m_{\rm bar}=(I/B_{\rm h})(2\pi/T)^2=1.01\unit{A\,m^2}$ for the bar, i.e.\ $m=0.252(12)\unit{A\,m^2}$ per cube, with an uncertainty of about $5\,\%$. Repelling magnets were held in place with calibrated non-magnetic spacers of known, predetermined thicknesses $s$ chosen to optimize the signal and configuration stability. Consequently, the geometrical parameters $d$, $z_2$, $x_2$, and $z_1$ in table~\ref{tab:moments} (and their corresponding prefactors $f$) are fully known beforehand for each assembled array.

The magnetic field was measured with the magnetometer of a Samsung Galaxy S10e smartphone using the phyphox app \citep{Staacks2018}. The device incorporates an AKM AK09918 triaxial magnetometer (developed by Asahi Kasei Microdevices), featuring a dynamic measurement range of $\pm 4900\unit{\mu T}$ in all three axes, a 16-bit ADC providing a digital resolution of $0.15\unit{\mu T/LSB}$, and an output data rate up to $100\unit{Hz}$. In our protocol, the app recorded the three components of the field during timed runs of $5\unit{s}$ (averaging over the recorded samples) and displayed their mean values and standard deviations; the standard deviation was taken as the statistical uncertainty of each reading, following \citet{Monteiro2021}. Before placing the magnets, the background field (the Earth's field plus the contribution of the surroundings and of the phone itself) was recorded in the same way and subtracted from every measurement.

\begin{figure}[tbp]
\centering
\includegraphics[width=0.49\textwidth]{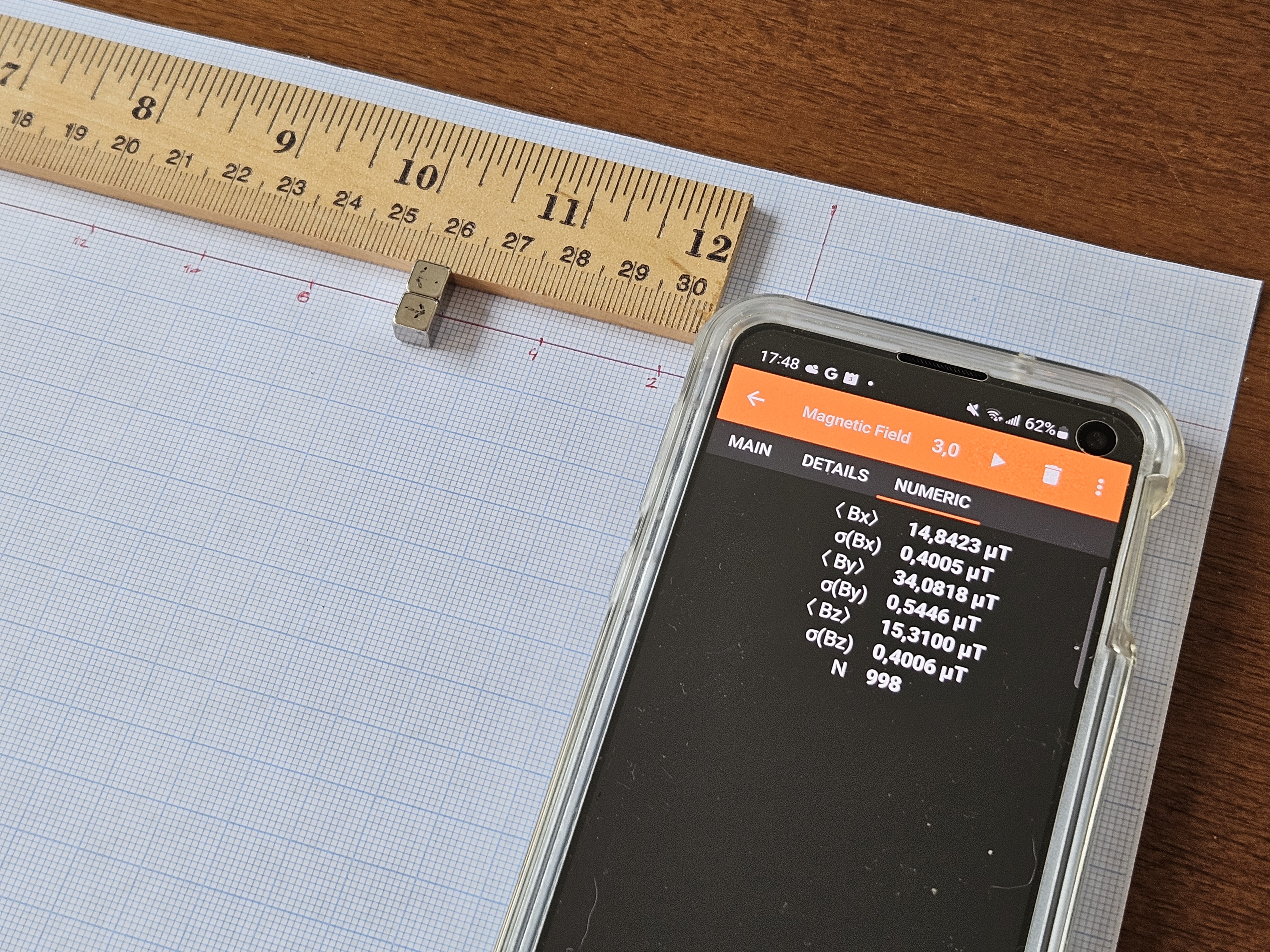}\hfill
\includegraphics[width=0.49\textwidth]{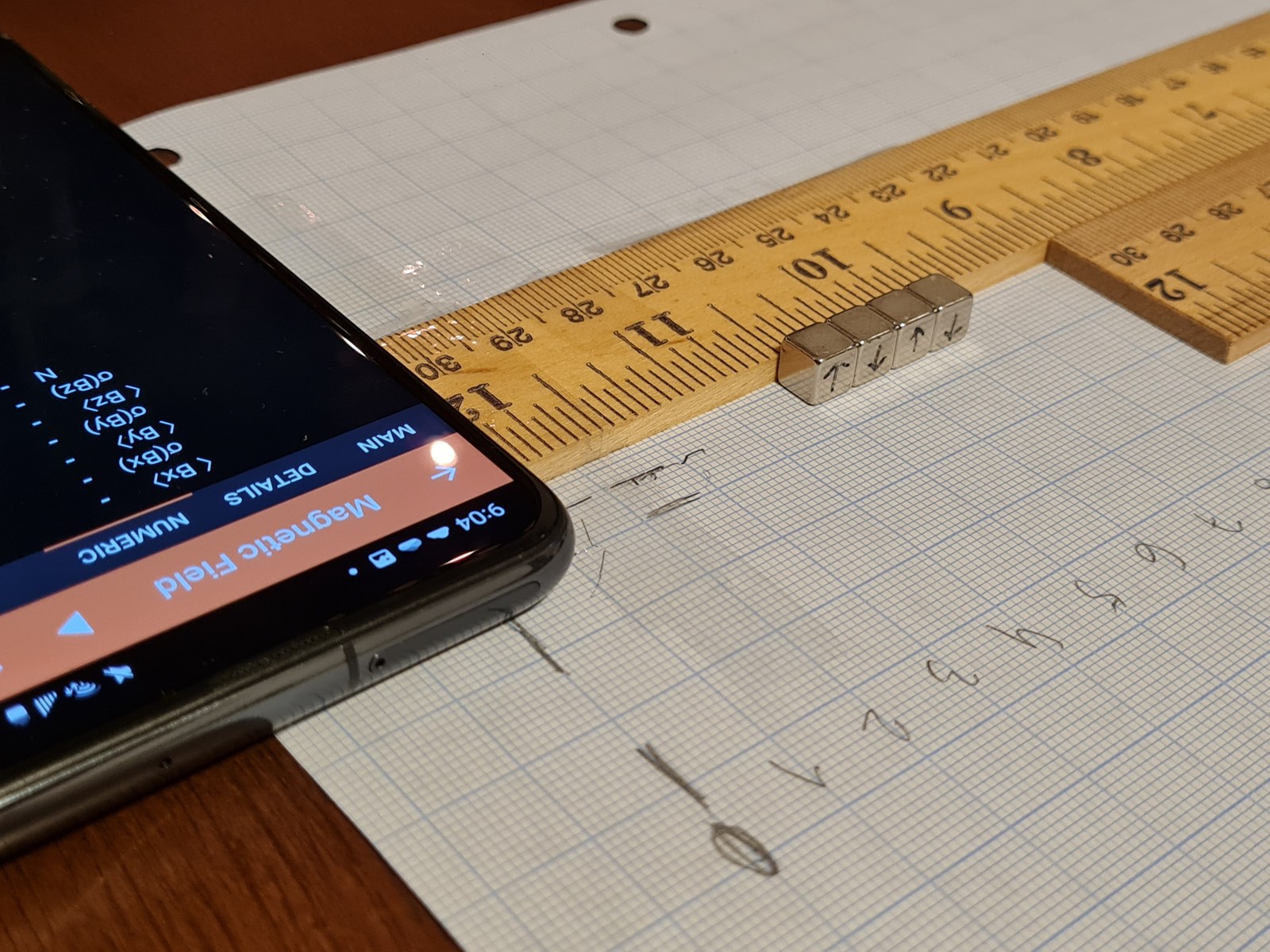}
\caption{Two examples of the experimental setup. According to the desired multipolar field, one, two, or four cube magnets are placed on a sheet of graph paper along the radial or axial axis of the smartphone magnetometer at different distances from the sensor. The app (phyphox) displays the mean value and the standard deviation of each component of the field during a timed run of $5\unit{s}$.}
\label{fig:setup}
\end{figure}

Two preliminary steps are essential. First, the position of the magnetometer chip inside the phone must be located, since all distances are measured from it; this is easily done by sliding a small magnet over the screen while watching the field reading, which is maximal when the magnet is directly above the sensor \citep{Arribas2015}. Second, the arrangement must be aligned with the sensor axes. The phone was laid flat on a sheet of graph paper with a ruler drawn along the $x$ axis of its sensor, passing through the magnetometer (figure~\ref{fig:setup}). For the radial measurements (\cfg{r}) the array was placed on this line with its magnetisation axis vertical, i.e.\ along the $z$ axis of the sensor: the tangential component is then simply the vertical component $B_z$ reported by the phone, and the polar component is $B_x$. For the axial measurements (\cfg{z}) the array was placed on the same line with its magnetisation axis along the line, pointing towards the sensor: the polar component is then $B_x$, and the tangential component is the component perpendicular to the line in the plane of the array ($B_y$ or $B_z$, depending on how the array is laid). The arrays were then moved along the ruler and the field recorded at distances between about $3\unit{cm}$ and $16\unit{cm}$ from the sensor. The lower limit is set by the requirement that the distance be large compared with the size of the array (and by the range of the sensor, of a few millitesla); the upper limit is reached when the field becomes comparable with the resolution and drift of the sensor, a few tenths of a microtesla, which for the octupoles happens at about $10$--$12\unit{cm}$.

\section{Measurements and results}
\label{sec:results}

For each configuration the magnitude $B$ of the surviving component, after subtraction of the background, was plotted against the distance $x$ in two complementary ways: as $\ln B$ versus $\ln x$, where a power law $B=C/x^n$ appears as a straight line of slope $-n$ and the exponent is obtained \emph{without assuming} its theoretical value, and as $B$ versus $1/x^n$ with the integer $n$ predicted by table~\ref{tab:moments}, where the slope gives directly the prefactor $C$ to be compared with the theory. The unweighted logarithmic regression yields the experimental decay exponent directly from the slope, whereas the linear regression weights closest points with higher absolute field values more heavily.

\begin{figure}[tbp]
\centering
\includegraphics[width=0.92\textwidth]{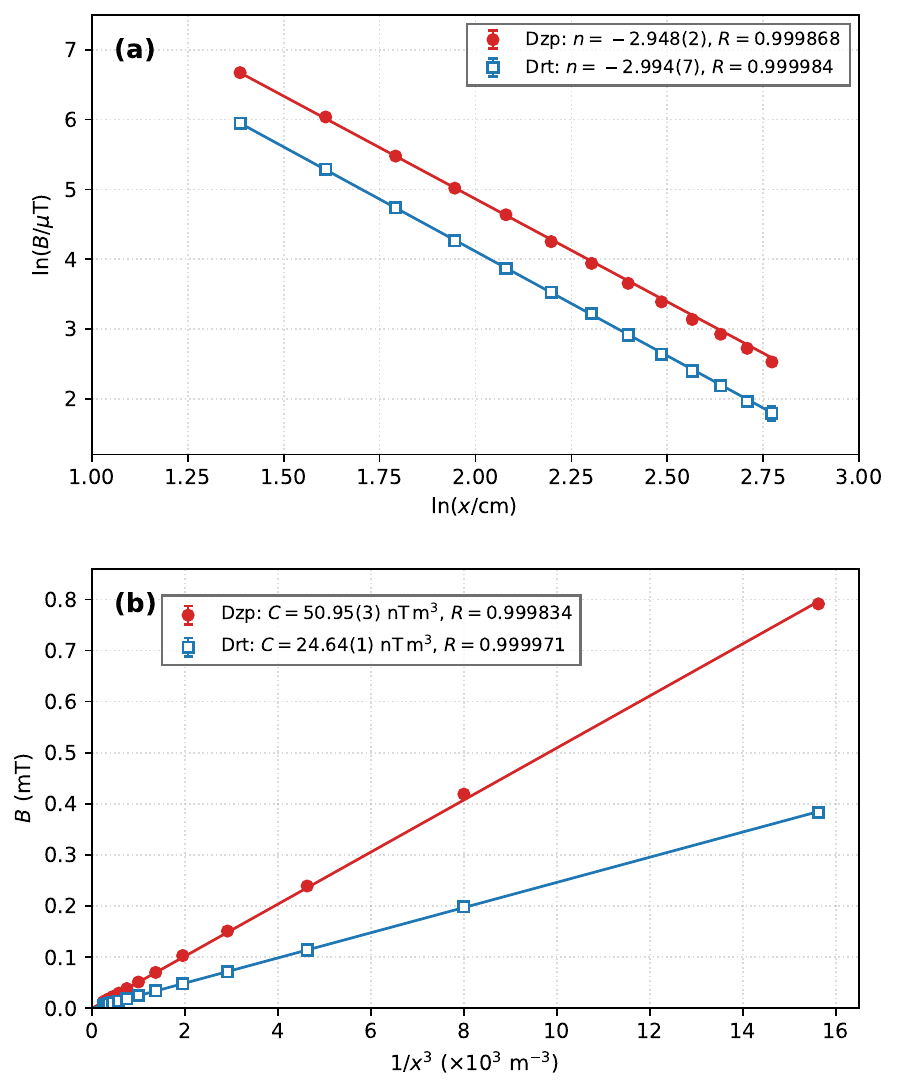}
\caption{Dipole. Top: $\ln B$ (with $B$ in $\mu$T) versus $\ln x$ (with $x$ in cm) for the polar (\cfg{Dzp}) and tangential (\cfg{Drt}) positions, with the linear fits and their fitted slopes $-n_{\mathrm{exp}}$. Bottom: the same data plotted against $1/x^3$; the slopes of the fits are the prefactors $C_{\mathrm{exp}}$ of table~\ref{tab:results}.}
\label{fig:dipole}
\end{figure}

\begin{figure}[tbp]
\centering
\includegraphics[width=0.92\textwidth]{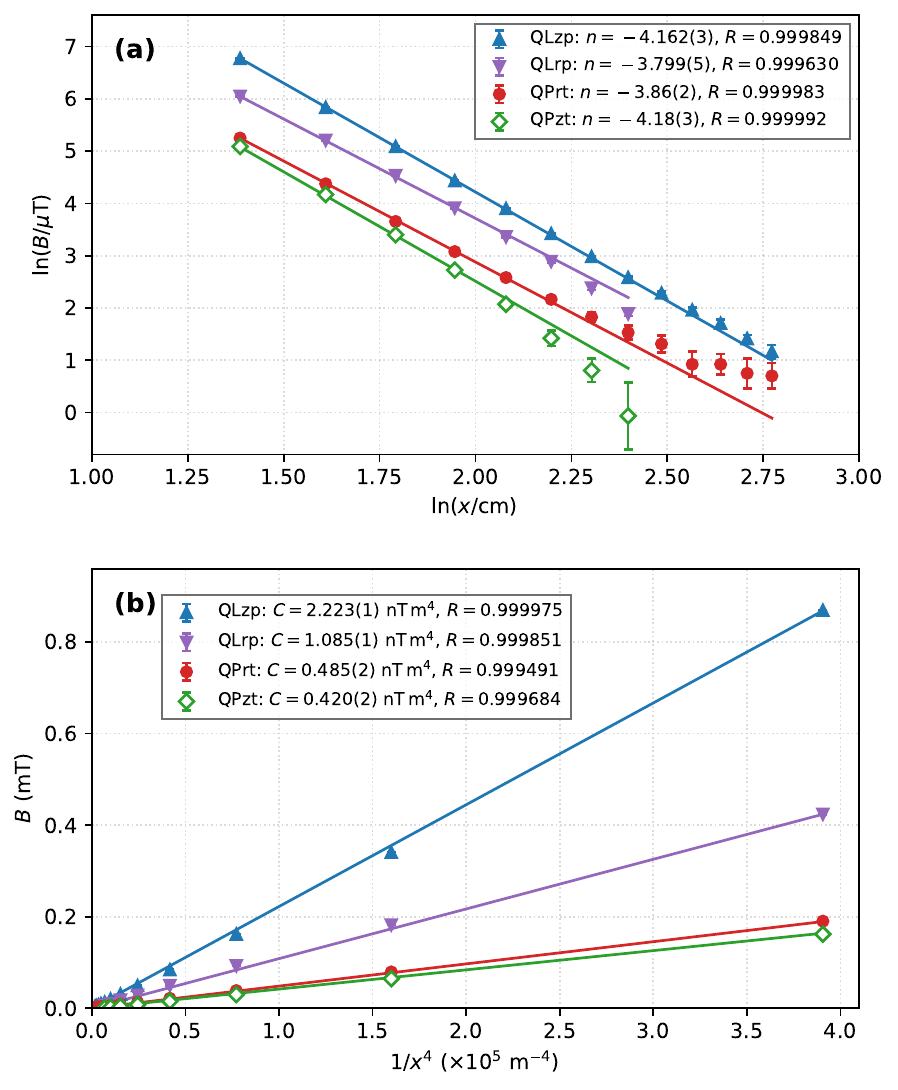}
\caption{Quadrupoles. Top: $\ln B$ versus $\ln x$ for the four quadrupole measurements (\cfg{QPrt}, \cfg{QPzt}, \cfg{QLzp}, \cfg{QLrp}) and their linear fits. Bottom: the same data plotted against $1/x^4$.}
\label{fig:quad}
\end{figure}

\begin{figure}[tbp]
\centering
\includegraphics[width=0.92\textwidth]{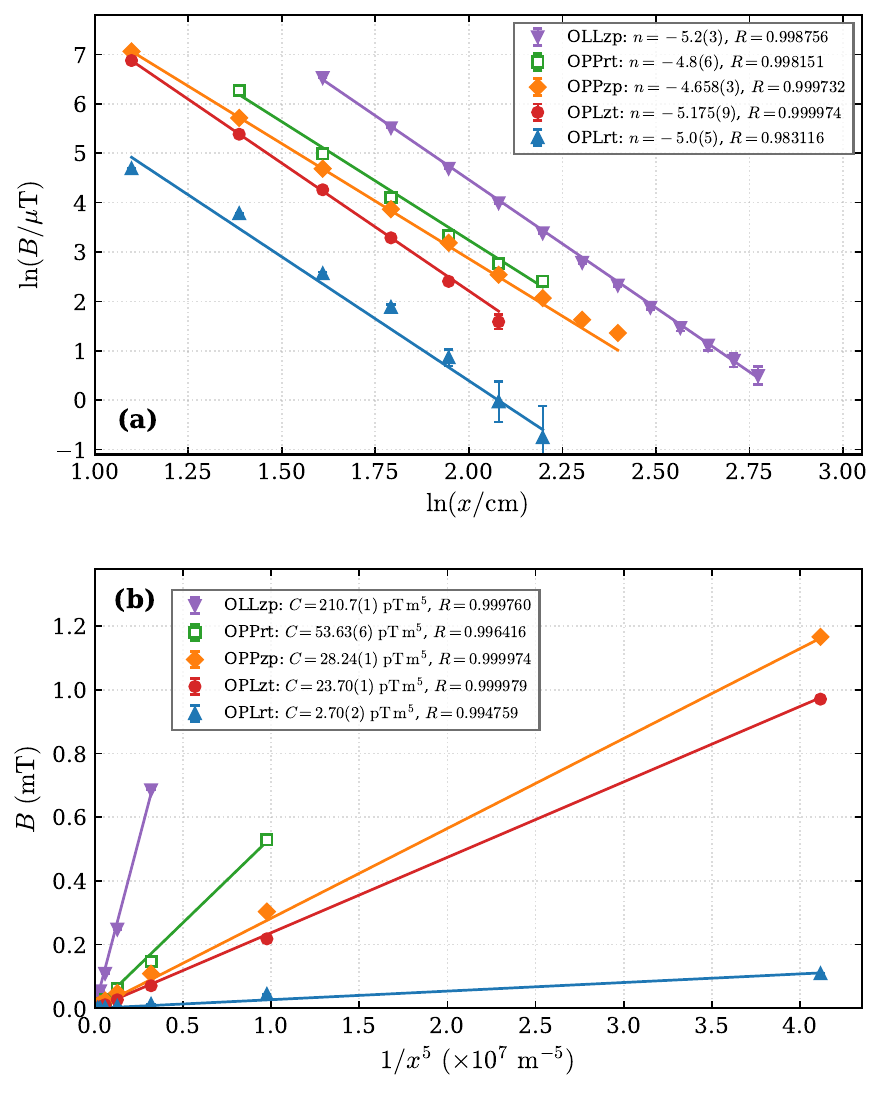}
\caption{Octupoles. Top: $\ln B$ versus $\ln x$ for the octupole measurements (\cfg{OPLzt}, \cfg{OPPrt}, \cfg{OLLzp}, \cfg{OPPzp}) and their linear fits; the series labelled \cfg{OPLrt} is the vertical component measured in the equatorial plane of the \cfg{OPL} array, which is expected to vanish by symmetry (see text). Bottom: the same data plotted against $1/x^5$.}
\label{fig:oct}
\end{figure}

Figure~\ref{fig:dipole} shows the results for the single magnet. The data follow straight lines over a factor of four in distance (from $\sim 4$ to $16\unit{cm}$) with slopes $-2.948(2)$ (\cfg{Dzp}) and $-2.994(7)$ (\cfg{Drt}), in excellent agreement with the expected inverse-cube law, and the prefactors, $50.95(3)$ and $24.64(1)\unit{nT\,m^3}$, agree with the predictions $2\kk m=50\unit{nT\,m^3}$ and $\kk m=25\unit{nT\,m^3}$ obtained with the independently measured $m=0.252(12)\unit{A\,m^2}$; their ratio, $2.068(2)$, reproduces the factor of 2 between the end-on and broadside fields of a dipole.

The four quadrupole measurements are shown in figure~\ref{fig:quad}. The fitted slopes $-n_{\mathrm{exp}}$ are $-3.86(2)$ (\cfg{QPrt}), $-4.18(3)$ (\cfg{QPzt}), $-4.162(3)$ (\cfg{QLzp}) and $-3.799(5)$ (\cfg{QLrp}), clustered around the expected value $-n_{\mathrm{th}}=-4$. The prefactors of the two \cfg{QP} measurements, $0.485(2)$ and $0.420(2)\unit{nT\,m^4}$, are both close to $3\kk ma=0.48\unit{nT\,m^4}$, confirming that for this arrangement the axial and radial positions give fields of comparable magnitude. For the \cfg{QL} quadrupole, assembled with a calibrated spacer $s=a$ ($d=2a$, giving $f=12$ and $f=6$), the prefactors $2.223(1)\unit{nT\,m^4}$ (axial) and $1.085(1)\unit{nT\,m^4}$ (radial) are in the ratio $2.05$, in agreement with the predicted factor of 2. Using these known geometric prefactors, the linearized fits directly yield the experimental dipole moments $m = C_{\mathrm{exp}}/(f\kappa a)$ reported in table~\ref{tab:results}: $0.2917(2)\unit{A\,m^2}$ and $0.2848(3)\unit{A\,m^2}$, respectively.

The octupole measurements are collected in figure~\ref{fig:oct}. The four configurations in which a non-vanishing component is predicted give slopes $-n_{\mathrm{exp}}$ of $-5.2(3)$ (\cfg{OLLzp}), $-4.658(3)$ (\cfg{OPPzp}), $-4.8(6)$ (\cfg{OPPrt}) and $-5.175(9)$ (\cfg{OPLzt}), clearly distinct from 3 and 4 and distributed around the expected value $-n_{\mathrm{th}}=-5$, with deviations of up to $0.34$. The spread is larger than for the dipole and the quadrupoles, for reasons discussed in the next section. The prefactors, $210.7(1)$, $28.24(1)$, $53.63(6)$ and $23.70(1)\unit{pT\,m^5}$ respectively, span an order of magnitude although the same four magnets are used in all cases; this directly reflects the sensitive dependence of the octupole moments on the array geometry and spacer dimensions (table~\ref{tab:moments}). Using the known prefactors $f$ determined by the calibrated spacers, we obtain from each linear fit the experimental dipole moment $m = C_{\mathrm{exp}} / (f\kappa a^2)$ (table~\ref{tab:results}). The resulting values of $m$ are consistent across all octupole configurations and show fair quantitative agreement (within $\sim 17\,\%$) with the independent value obtained from the pendulum oscillation, with remaining differences attributable to finite-size corrections and small positioning tolerances.

Figure~\ref{fig:oct} also includes the vertical component measured in the equatorial plane of the \cfg{OPL} array (series \cfg{OPLrt}). According to equation~(\ref{eq:radial_t}) this component vanishes identically for the point-dipole model, since $M_{00}=M_{10}=M_{20}=M_{02}=0$ for this array: the surviving field in the equatorial plane is polar (\cfg{OPLrp}, table~\ref{tab:moments}). The measured signal is indeed almost an order of magnitude smaller than the other octupole fields at the same distance. It is not exactly zero: a slight angular tilt $\theta$ between the magnetisation axis and the sensor axes mixes a fraction of the much larger polar component into the vertical reading ($B_{\mathrm{res}} \sim \theta B_{\mathrm{polar}}$), which naturally preserves the $x^{-5}$ decay (the fitted slope is $-5.0(5)$). Small inequalities among individual magnet moments may also leave residual dipole ($x^{-3}$) or quadrupole ($x^{-4}$) contributions, but the observed steep slope indicates that geometrical axis misalignment is the dominant contribution here.

\begin{table}[tbp]
\centering
\caption{Summary of the measurements (figures~\ref{fig:dipole}--\ref{fig:oct}) compared with the leading-order predictions ($|B|=f\kk ma^{n-3}/x^n$, with $a=0.635\unit{cm}$ and independent reference $m=0.252(12)\unit{A\,m^2}$). For each configuration and geometric prefactor $f$ (table~\ref{tab:moments}), the table lists the theoretical and experimental slopes with $|R_{\mathrm{log}}|$ from the $\ln B$ vs.\ $\ln x$ regression, and the fitted prefactor $C_{\mathrm{exp}}$, inferred single-magnet moment $m=C_{\mathrm{exp}}/(f\kk a^{n-3})$ and $|R_{\mathrm{lin}}|$ from the linear $B$ vs.\ $1/x^n$ regression. Parentheses denote the statistical regression uncertainty in the last digit.}
\label{tab:results}
\vspace{0.5ex}
\footnotesize
\setlength{\tabcolsep}{3.2pt}
\begin{tabular}{@{}lccccccc@{}}
\toprule
 & & \multicolumn{3}{c}{Log-log fit ($\ln B$ vs.\ $\ln x$)} & \multicolumn{3}{c}{Linearized fit ($B$ vs.\ $1/x^n$)}\\
\cmidrule(lr){3-5}\cmidrule(lr){6-8}
Code & $f$ & Slope $-n_{\mathrm{th}}$ & Fit slope $-n_{\mathrm{exp}}$ & $|R_{\mathrm{log}}|$ & $C_{\mathrm{exp}}$ & $m$ ($\mathrm{A\,m^2}$) & $|R_{\mathrm{lin}}|$\\
\midrule
\multicolumn{8}{@{}l}{\emph{Dipole fields} ($n_{\mathrm{th}}=3$, $C_{\mathrm{exp}}$ in $\mathrm{nT\,m^3}$)}\\
\cfg{Dzp}   & 2            & $-3$ & $-2.948(2)$ & $0.999868$ & $50.95(3)$ & $0.2548(2)$ & $0.999834$\\
\cfg{Drt}   & 1            & $-3$ & $-2.994(7)$ & $0.999984$ & $24.64(1)$ & $0.2464(1)$ & $0.999971$\\
\midrule
\multicolumn{8}{@{}l}{\emph{Quadrupole fields} ($n_{\mathrm{th}}=4$, $C_{\mathrm{exp}}$ in $\mathrm{nT\,m^4}$)}\\
\cfg{QLzp}  & $12^{\rm a}$ & $-4$ & $-4.162(3)$ & $0.999849$ & $2.223(1)$ & $0.2917(2)$ & $0.999975$\\
\cfg{QLrp}  & $6^{\rm a}$  & $-4$ & $-3.799(5)$ & $0.999630$ & $1.085(1)$ & $0.2848(3)$ & $0.999851$\\
\cfg{QPzt}  & 3            & $-4$ & $-4.18(3)\phantom{0}$ & $0.999992$ & $0.420(2)$ & $0.220(1)\phantom{0}$ & $0.999684$\\
\cfg{QPrt}  & 3            & $-4$ & $-3.86(2)\phantom{0}$ & $0.999983$ & $0.485(2)$ & $0.255(1)\phantom{0}$ & $0.999491$\\
\midrule
\multicolumn{8}{@{}l}{\emph{Octupole fields} ($n_{\mathrm{th}}=5$, $C_{\mathrm{exp}}$ in $\mathrm{pT\,m^5}$)}\\
\cfg{OLLzp} & $192^{\rm b}$& $-5$ & $-5.2(3)\phantom{00}$ & $0.998756$ & $210.7(1)\phantom{0}$ & $0.2722(2)$ & $0.999760$\\
\cfg{OPLzt} & $24^{\rm c}$ & $-5$ & $-5.175(9)$ & $0.999974$ & $23.70(1)$ & $0.2449(1)$ & $0.999979$\\
\cfg{OPLrt} & 0            & ---  & $-5.0(5)\phantom{00}$ & $0.983116$ & $2.70(2)$  & ---         & $0.994759$\\
\cfg{OPPzp} & $24^{\rm d}$ & $-5$ & $-4.658(3)$ & $0.999732$ & $28.24(1)$ & $0.2918(1)$ & $0.999974$\\
\cfg{OPPrt} & $48^{\rm e}$ & $-5$ & $-4.8(6)\phantom{00}$ & $0.998151$ & $53.63(6)$ & $0.2771(3)$ & $0.996416$\\
\bottomrule
\end{tabular}
\par\smallskip
\begin{minipage}{\linewidth}\footnotesize\raggedright
$^{\rm a}$ Nominal value for $d=2a$ ($s=a$).\par
$^{\rm b}$ Value for $z_2^2-z_1^2=8a^2$ (\cfg{OLLzp}) with spacer $s\simeq 1.37a$.\par
$^{\rm c}$ Value for nominal spacer $s=a$ ($x_1=a/2, z_1=a$).\par
$^{\rm d}$ Value for touching inner magnets ($s=0$, $x_2^2-x_1^2=2a^2$).\par
$^{\rm e}$ Value for nominal spacer $s=2a$ ($x_2^2-x_1^2=4a^2$).
\end{minipage}
\end{table}

\section{Discussion}
\label{sec:discussion}

The central result of the experiment is that, with the same four magnets and the same phone, students measure three clearly different decay laws, $1/x^3$, $1/x^4$ and $1/x^5$, and can trace each of them to the cancellation of the lower-order moments in table~\ref{tab:moments}. The exponents are determined to within a few hundredths for the dipole, a few tenths for the quadrupoles and up to $0.34$ for the octupoles. The uncertainties reported in table~\ref{tab:results} reflect the statistical standard error of the linear fits; the fact that deviations from theoretical integers exceed these statistical margins illustrates the presence of systematic physical effects governing higher-order multipoles.

\emph{Higher-order terms at short distance.} Equations~(\ref{eq:axial_p})--(\ref{eq:radial_p}) are the first terms of series in the ratio of source size to observation distance, and the relative weight of the next term grows with the order of the multipole. For the \cfg{OPP} octupole measured along the $x$ axis, for instance, the next term in equation~(\ref{eq:radial_t}) is $-15\kk M_{40}/x^7$ with $M_{40}=2m(x_2^4-x_1^4)$, so that the relative correction is $\frac{5}{2}(x_1^2+x_2^2)/x^2$; with the inner magnets touching ($s=0$) this equals $\frac{25}{4}(a/x)^2$, i.e.\ $28\,\%$ at $x=3\unit{cm}$ and $10\,\%$ at $x=5\unit{cm}$. Positive corrections of this kind make the apparent local exponent $-{\rm d}\ln B/{\rm d}\ln x$ larger than 5 at the near end of the range, and they are strongest for the arrangements with the largest $x_2$ or $z_2$, which is consistent with the slopes steeper than 5 found for \cfg{OLLzp} and \cfg{OPLzt}. In addition, the finite size of the cube magnets introduces corrections beyond the point-dipole approximation, although for an ideal uniformly magnetised cube the cubic symmetry cancels internal quadrupole corrections, shifting the leading intrinsic term to order $(a/R)^4$ \citep{Binder2016}.

\emph{Residual lower-order moments at long distance.} Conversely, any imperfection in the cancellation of the dipole and quadrupole moments leaves residual fields that decay more slowly than the octupole and eventually dominate it. If one of the four magnets has a moment that differs by a fraction $\epsilon$ from the others, the array has a residual dipole moment $\epsilon m$ whose equatorial field, $\kk\epsilon m/x^3$, must be compared with the octupole field $24\kk ma^2/x^5$ of a touching \cfg{OPP} array: the ratio is $\epsilon x^2/24a^2$, which for $\epsilon=0.05$ reaches $0.5$ at $x=10\unit{cm}$. A displacement of one magnet by $\delta$ along the line produces a residual quadrupole moment $m\delta$ whose field, $3\kk m\delta/x^4$, is a fraction $\delta x/8a^2$ of the octupole field; for $\delta=0.5\unit{mm}$ this is $15\,\%$ at $x=10\unit{cm}$. Depending on the relative sign, residual moments that reinforce the octupole field reduce the apparent exponent below 5, whereas opposing residual moments tend to increase it or cause local cancellations. In figure~\ref{fig:oct}, upward curvature is visible in \cfg{OPPzp} and \cfg{OPPrt} beyond $\ln x\simeq 2.2$ ($x\simeq 9\unit{cm}$), where the signal also approaches the noise floor of the sensor. Octupole fields are thus `squeezed' from both sides, and the useful window---roughly from $4a$ to $15a$ here---is much narrower than for the dipole.

\emph{Location of the sensor.} All distances are measured from the magnetometer chip, whose position inside the phone is known to about $\pm 2\unit{mm}$ \citep{Arribas2015}. A systematic error $\delta x$ in this position changes the fitted exponent of a pure power law by approximately $n\,\delta x/x$, which for $n=5$ and $\delta x = \pm 2\unit{mm}$ evaluates to $\pm 0.20$--$0.33$ across $x=3$--$5\unit{cm}$. This is probably the dominant systematic effect for configurations measured in the axial position, where the sensor and array planes must be carefully aligned. A refinement accessible to advanced students is to fit the data with the exact sum of point-dipole fields leaving sensor offset and spacer thickness as adjustable parameters, or using open-source packages such as Magpylib \citep{Magpylib2020,Sullivan2023}.

\emph{Prefactors and the moment of the magnets.} The comparison between experimental prefactors and theoretical predictions provides a direct, quantitative consistency check across all configurations. For the spacer-free arrangements (\cfg{Dzp}, \cfg{Drt}, \cfg{QPzt}, \cfg{QPrt}), the experimental prefactors agree with the predictions based on the independent pendulum reference $m=0.252(12)\unit{A\,m^2}$ within $2\,\%$ to $12\,\%$. For arrangements with spacers, since the geometry and spacer dimensions $s$ are known a priori from the design of the arrays, the dimensionless factor $f$ allows us to systematically determine the experimental dipole moment $m = C_{\mathrm{exp}}/(f\kk a^{n-3})$ (table~\ref{tab:results}). The fair consistency of the extracted values of $m$ (within $\sim 17\,\%$ across all multipole orders under nominal geometry) demonstrates the quantitative predictive power of the unified model. Furthermore, the factor-of-two difference between the two \cfg{OPP} prefactors ($53.63$ vs.\ $28.24\unit{pT\,m^5}$) directly confirms the spacer dependence: since $x_2^2-x_1^2 = a(2a+s)$, the prefactor ratio is $C(s)/C(0) = 1 + s/(2a)$, so that inserting a known spacer of two magnet widths ($s=2a$) doubles the resulting field.

\emph{Which octupole?} The three octupoles of figure~\ref{fig:config} are not equivalent: they possess different non-vanishing moments ($M_{02}$, $M_{20}$ and $M_{11}$), producing fields of distinct symmetry and magnitude. The most favourable configuration for laboratory detection is \cfg{OLL}, whose moment increases rapidly with outer magnet separation and whose axial polar field has the largest numerical coefficient in equation~(\ref{eq:axial_p}).

\emph{Pedagogical remarks.} The activity can be organised as a sequence of increasing difficulty that fits a single three-hour laboratory session or a short project. Students first measure the dipole, locating the sensor and learning to subtract the background; then they build the two quadrupoles, predict with table~\ref{tab:moments} which component survives at each position, and check the $1/x^4$ law; finally they build one or more octupoles. The planning stage---deciding which component to read, how to align the array and over which range of distances the measurement is meaningful---is where most learning takes place, and the `trap' of the alternating $+\,-\,+\,-$ arrangement, whose $1/x^4$ law the phone reveals immediately, provides a clear demonstration that cancelling the dipole moment is not sufficient to obtain an octupole.

\section{Conclusions}
\label{sec:conclusions}

We have shown that the multipole expansion of the magnetic field can be explored quantitatively in an introductory laboratory with four identical cube magnets and the magnetometer of a smartphone. A simple notation, based on how the magnets are joined (longitudinally or in parallel), where the sensor is placed (on the axis or in the equatorial plane) and which component is read (polar or tangential), together with the spatial Cartesian moments $M_{pq}$ of the array and the four expansions (\ref{eq:axial_p})--(\ref{eq:radial_p}), allows students to design dipole, quadrupole and octupole arrangements, to predict from symmetry which component survives, and to compare measured decay laws and prefactors with theory. The measured exponents, $n\simeq3$, $4$ and $5$, and the consistency of the prefactors with the independently determined dipole moment of the magnets confirm the validity of the unified model. The analysis of deviations---higher-order terms at short distance, residual lower-order moments at long distance, and sensor placement uncertainty---gives students a realistic view of experimental multipole characterisation.

From a broader perspective, the methodology presented here transcends its pedagogical value, offering a robust framework for low-cost magnetic instrumentation. The capacity to experimentally isolate specific multipole contributions and verify analytical decay laws using ubiquitous hardware provides researchers and engineers with an accessible tool for the rapid prototyping, testing, and field characterization of customized permanent-magnet structures without relying on expensive laboratory equipment.

\section*{Acknowledgements}
M.M., C.S. and A.C.M. acknowledge support from \textit{Programa de Desarrollo de las Ciencias B\'asicas} (UdelaR, MEC, Uruguay).
E.A., I.E. and R.R.V. gratefully acknowledge financial support from the Junta de Comunidades de Castilla-La Mancha of Spain (Project SBPLY/\allowbreak23/\allowbreak180225/\allowbreak00089). A.B. gratefully acknowledges financial support from the Generalitat Valenciana of Spain (project CIPROM/2024/90) and the Ministry of Science, Innovation and Universities of Spain (project PID2024-161610OB-I00).


\end{document}